\documentclass[aps,prl,twocolumn,groupedaddress]{revtex4}  
\usepackage{graphicx}  
\usepackage{dcolumn}   
\usepackage{bm}        
\usepackage{amssymb}   
\usepackage{amsmath}   
\usepackage{mathtools} 
\usepackage{xcolor}
\usepackage{ulem}

\usepackage[colorlinks,linkcolor=blue,urlcolor=blue,citecolor=blue]{hyperref}

\begin{document}

\title{Effective Resource-Competition Model for Species Coexistence} 
\author{Deepak Gupta$^{1,2}$}
\thanks{These authors equally contributed to this work}
\author{Stefano Garlaschi$^{1}$}
\thanks{These authors equally contributed to this work}
\author{Samir Suweis$^{1}$}
\author{Sandro Azaele$^{1}$}
\thanks{Equal senior contribution}
\author{Amos Maritan$^{1}$}
\thanks{Equal senior contribution}
\affiliation{$^1$Dipartimento di Fisica e Astronomia “Galileo Galilei”, Universit\`a degli Studi di Padova, via Marzolo 8, 35131 Padova, Italy}
\affiliation{$^2$Department of Physics, Simon Fraser University, Burnaby, British Columbia V5A1S6, Canada}
\date{\today}

\begin{abstract}
Local coexistence of species in large ecosystems is traditionally explained within the broad framework of niche theory. However, its rationale hardly justifies rich biodiversity observed in nearly homogeneous environments. Here we consider a consumer-resource model in which a coarse-graining procedure accounts for a variety of ecological mechanisms and leads to effective spatial effects which favour species coexistence. Herein, we provide conditions for several species to live in an environment with very few resources. In fact, the model displays two different phases depending on whether the number of surviving species is larger or smaller than the number of resources. We obtain conditions whereby a species can successfully colonize a pool of coexisting species. Finally, we analytically compute the distribution of the population sizes of coexisting species. Numerical simulations as well as empirical distributions of population sizes support our analytical findings.

\end{abstract}

\maketitle
Our planet hosts an enormous number of species \cite{Book-1}, which thrive within a variety of environmental conditions. The coexistence of this enormous biological diversity is traditionally explained in terms of local adaptation \cite{thompson1994coevolutionary,schluter2000ecology}, environmental heterogeneity \cite{chesson1997roles,chesson2000mechanisms}, species' abilities to aptly respond to the distribution of resources \cite{vincent1996trade,schmidt2010ecology}, and other abiotic factors which broadly define a niche \cite{chase2003ecological}.
When species are geographically separated, they may survive because they match a specific environmental condition and inter-specific competition is not detrimental. However, several microbial species seem to coexist despite they occupy very similar niches in close-by regions \cite{paradox-2,Zelezniak6449,Goldford469}. 
This scenario is known as the {\it paradox of plankton} \cite{paradox}. Now, on timescales that are larger than one generation but smaller than speciation timescales, the fittest species should outcompete all the others. Then, why do we still observe coexistence? Consistent with this rationale, theoretical work based on MacArthur's {\it consumer-resource model} \cite{Arthur,Chesson-MA} confirms the validity of the so called {\it competitive exclusion principle} (CEP): the number of coexisting species competing for the same resources is bounded by the number of resources themselves \cite{exclusion-1,exclusion-2,exclusion-3,mcgehee1977some,armstrong1980competitive,britton1989aggregation,hening2020competitive}. Despite numerous attempts \cite{Wingreen,Leo,Roy-2}, no definitive answer has yet been achieved for explaining such stark contrast between the predictions of CEP and species' coexistence.

A complementary framework, which has been able to explain several biodiversity patterns at macroscopic scales, is the neutral theory of biogeography \cite{caswell1976community,hubbell1979tree,hubbell2001unified,vallade2003analytical,chave2004neutral,alonso2006merits,rosindell2011unified,azaele2016statistical}. 
Instead of looking at what specific traits facilitate species' survival, this approach highlights the general features which tend to make species more similar to each other. This framework has the merit to predict several patterns in agreement with the empirical data \cite{hubbell1997unified,volkov2003neutral,azaele2006dynamical,peruzzo2020spatial,bell2000distribution,bell2001neutral}. 
However, it lacks a convincing mechanism of coexistence, which is usually maintained only by an external source of individuals. Within this context, competition for resources plays a relatively minor role with respect to the niche setting, and the total number of species sustained by a region cannot be inferred by the availability of resources in the habitat. 
 
In this Letter, we propose an alternative approach based on a generalization of the aforementioned MacArthur's consumer-resource model. This new formulation explains why
a large number of species can coexist even in the presence of a limited number of resources, thus violating CEP. Secondly, it predicts how many species will survive depending on the amount of resource present in the habitat. Thirdly, we find the conditions under which an invading species outcompetes a pool of coexisting species. Finally, we analytically obtain the species abundance distribution (SAD), i.e., the probability distribution of the population sizes of the species, and show that it justifies the empirical SAD calculated from the plankton data presented in Ref.~\cite{data-1}.
 
The key feature of this new framework is the emergence of new terms which stabilize species interactions and affect the dynamics on top of the traditional inter-species couplings, which account for the indirect resource consumption. These stabilizing factors emerge naturally in all ecosystems when spatial effects are not negligible. Indeed, by 
coarse-graining 
the spatial degrees of freedom, we show that a density-dependent inhibition term forms and stabilizes the dynamics.
For example in tree communities, 
this term may model the \textit{Janzen-Connell effect} (JCE) \cite{janzen1970herbivores,connell1971role,J-C}, that describes the inhospitability for the seedlings in the proximity of parent trees due to host-specific pathogens  (in the SM we provide a simple derivation of this result).
This leads to a penalization of their growth and 
inhibits the local crowding of individuals belonging to the same species \cite{clark1984spacing,bagchi2010testing,bagchi2014pathogens}.

We consider an ecological community composed of $M$ different species competing for $R$ resources. These species are characterized by their maximum consumption rates, $\alpha_{\sigma i}$, at which a species $\sigma$ uptakes the $i$-th resource and converts it into its biomass at high concentration of the resource, $c_i$. In what follows, we also refer to the $\alpha$'s as  {\it metabolic strategies}. Since per-capita growth rates are proportional to the resource concentrations when they are low,  the overall dependence of resource concentration is typically captured by multiplying the maximum growth rate by the {\it Monod function}, $r_i(c_i)=c_i/(k_i+c_i)$, where $k_i$ is a resource dependent constant. Further, resources degrade in time with a rate $\mu_i$. 
Similarly, populations decay with intrinsic mortality rate $\beta_\sigma$. Therefore, the system evolves with a spatially-extended consumer-resource model~\cite{Arthur,Chesson-MA}:
\begin{align}
    \dot{n}_\sigma(\vec x)&=n_\sigma(\vec x)\left[\sum_{i=1}^R\alpha_{\sigma i} r_i\left(c_i(\vec x)\right)-\beta_\sigma\right]-\vec\nabla \cdot J_\sigma(\vec x,t),\label{sp-eqn-space}\\
  \dot c_i(\vec x)&=\mu_i(\Lambda_i-c_i(\vec x))-r_i(c_i(\vec x))\sum_{\sigma=1}^M n_\sigma(\vec x) \alpha_{\sigma i}, \label{res-eqn-space}
\end{align}
where $\vec x$ indicates spatial degrees of freedom and the flux $J_\sigma(\vec x,t)$ originates from a variety of spatial ecological mechanisms, including foraging/chemotaxis \cite{KELLER1971225}, crowding effects of species competing for resources in limited areas or species-induced modification of the environment for a competitive advantage \cite{garcia2018bacterial,granato2019evolution,giometto2020antagonism}. The approach is not limited to a specific spatial effect. Performing the spatial coarse-graining \cite{RG-1,RG-2} [see 
\cite{SM} for $J_\sigma(\vec x,t)$ and detailed derivation on coarse-graining], we eventually end up with the following effective consumer-resource model:
\begin{align}
  \dot{n}_\sigma&=n_\sigma\big[\sum_{i=1}^R\alpha_{\sigma i} r_i(c_i)-\beta_\sigma-\sum_{\rho=1}^{M}\epsilon_{\sigma\rho} n_\rho\big],\label{sp-eqn-3}\\
  \dot c_i&=\mu_i(\Lambda_i-c_i)-r_i(c_i)\sum_{\sigma=1}^M n_\sigma \alpha_{\sigma i}, \label{res-eqn-3}
\end{align}
where the term  $\epsilon_{\sigma \rho}$ in Eq.~\eqref{sp-eqn-3} is the interaction of species $\rho$ with species $\sigma$ because of the spatial coarse-graining. For some ecological mechanisms the parameters $\epsilon_{\sigma \rho}$ and $\alpha_{\sigma i}$ may be correlated \cite{SM}. For instance, if $\epsilon_{\sigma \rho}$ are determined by resource-specific foraging alone, they depend on the metabolic strategies $\alpha_{\sigma i}$ and in this case there may be no increase in biodiversity relative to a
non-spatial model. However, for other mechanisms the two matrices are independent, as for the case of crowding of species or the presence of pathogens \cite{SM}, which can also generate the last term in Eq.~\eqref{sp-eqn-3}. In the following we will focus on the case of independent matrices.
Here $\epsilon_{\sigma \sigma}^{-1}$ may be treated as proportional to the carrying capacity for the species $\sigma$. In Eq.~\eqref{res-eqn-3}, the quantity $\mu_i\Lambda_i$ is the rate of supplying {\it abiotic} resources. Notice that biotic resources are typically modeled by substituting $\mu_i$ with $\mu_ic_i$.
Then, $\mu_i\Lambda_i$ and $\Lambda_i$, respectively, correspond to the growth rate and the carrying capacity of the $i$-th resource.
In what follows, for simplicity, we report the results for the case of abiotic resources  with degradation rates of all resources to be the same: $\mu=\mu_i$. The other cases do not display qualitatively different results.
\begin{figure}
    \centering
    \includegraphics[width=4.3cm]{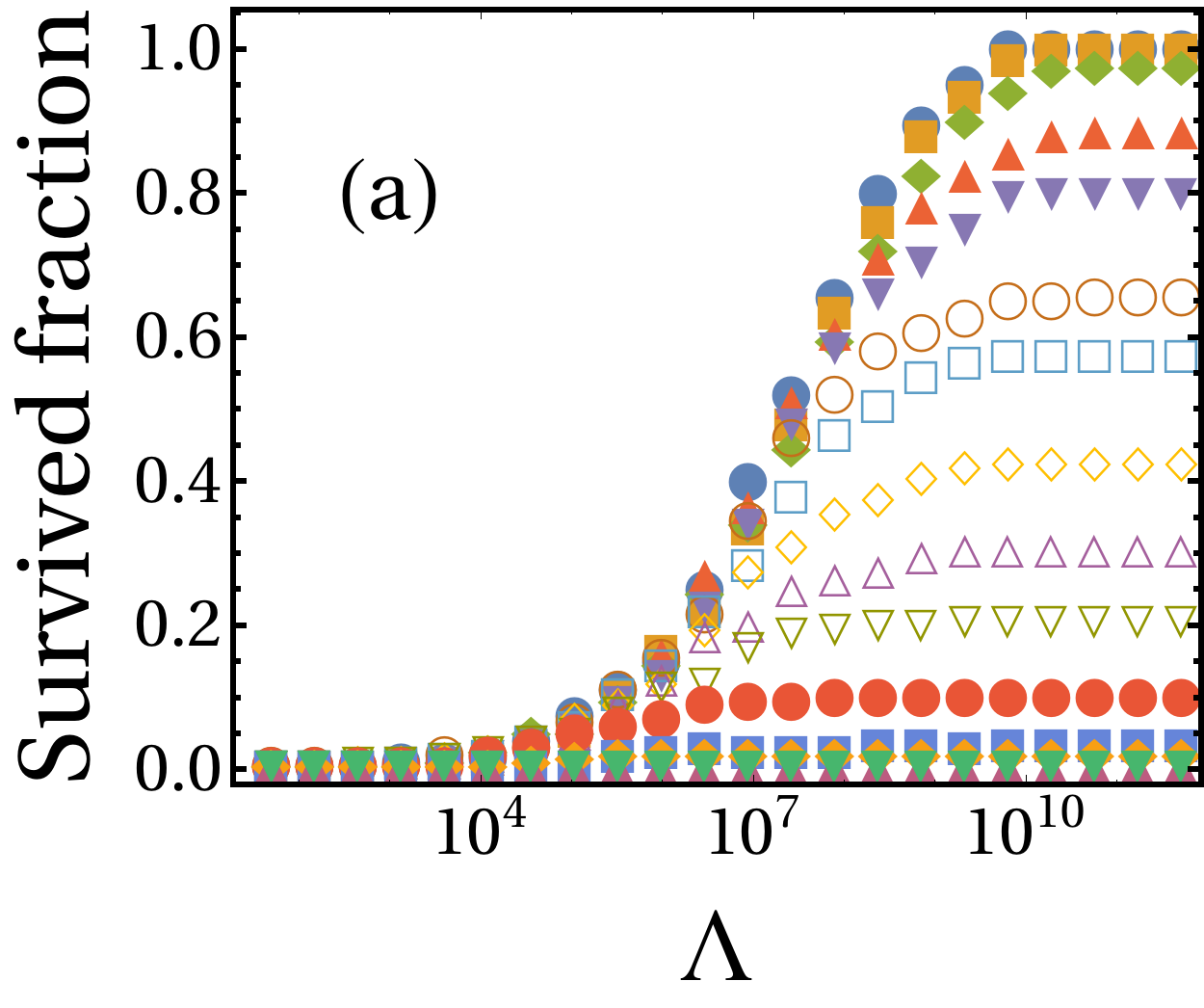}~
    \includegraphics[width=4.3cm]{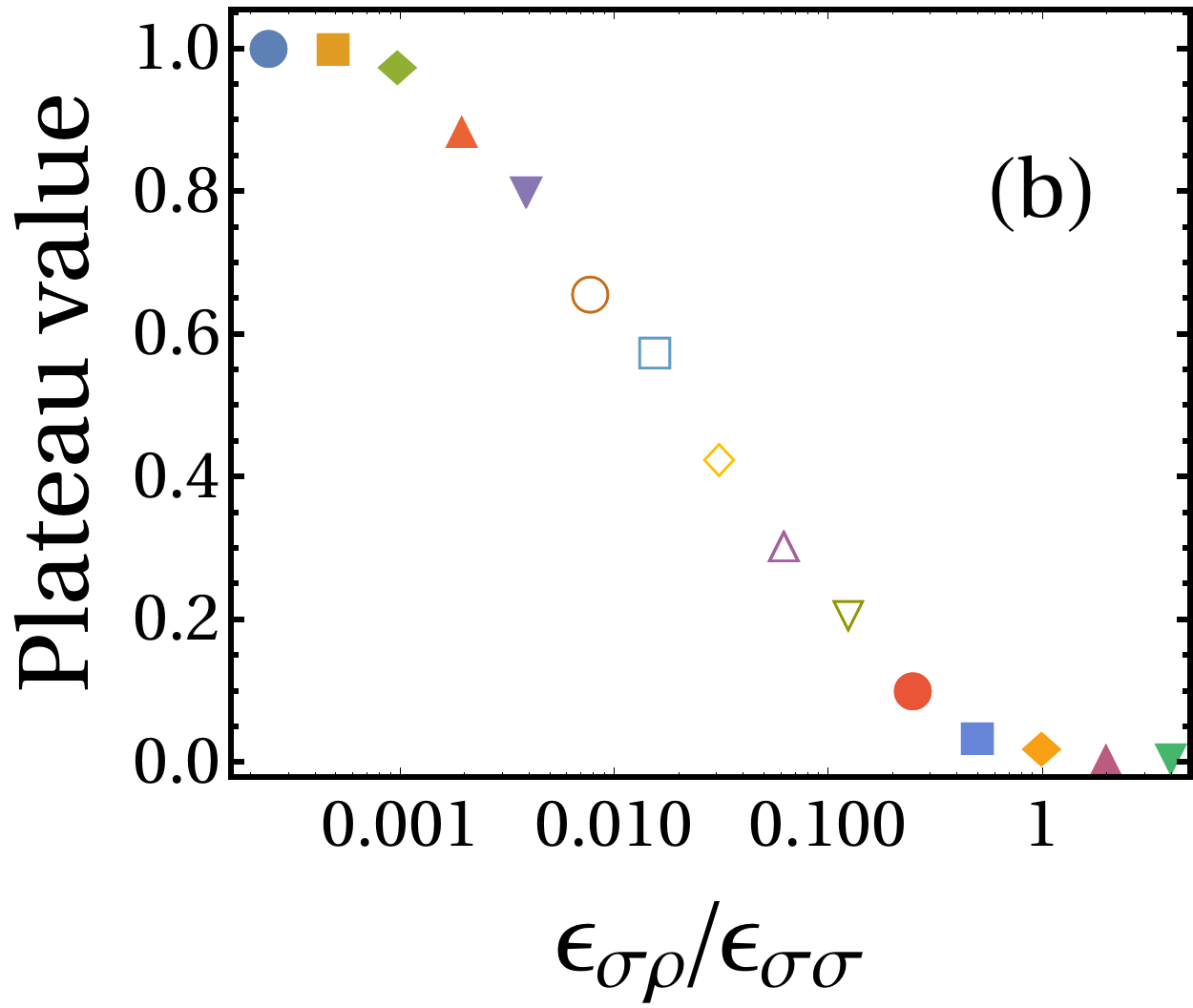}~~
    \caption{Fraction of survived species. Panel (a): We numerically evolve dynamics \eqref{sp-eqn-3} and \eqref{res-eqn-3} up to time $t=10^{30}$ and compute the fraction of survived species as a function of $\Lambda$. The ratio of inter- to intra-species interaction, i.e., $\epsilon_{\sigma \rho}/\epsilon_{\sigma \sigma}$, is increased by a power of 2 as we go from the top saturating curve ($\epsilon_{\sigma \rho}/\epsilon_{\sigma \sigma}=2^{-12}$) to the bottom one ($\epsilon_{\sigma \rho}/\epsilon_{\sigma \sigma}=2^{2}$). Here we take the initial number of species $M=200$ and $R=1$. Panel (b): Plot for the corresponding plateau points of panel(a) (with same color coding) as a function of the ratio $\epsilon_{\sigma \rho}/\epsilon_{\sigma \sigma}$. For details see \cite{SM}.}
    \label{fig:non-diag-E-species}
\end{figure}
\begin{figure}
    \includegraphics[width=\columnwidth]{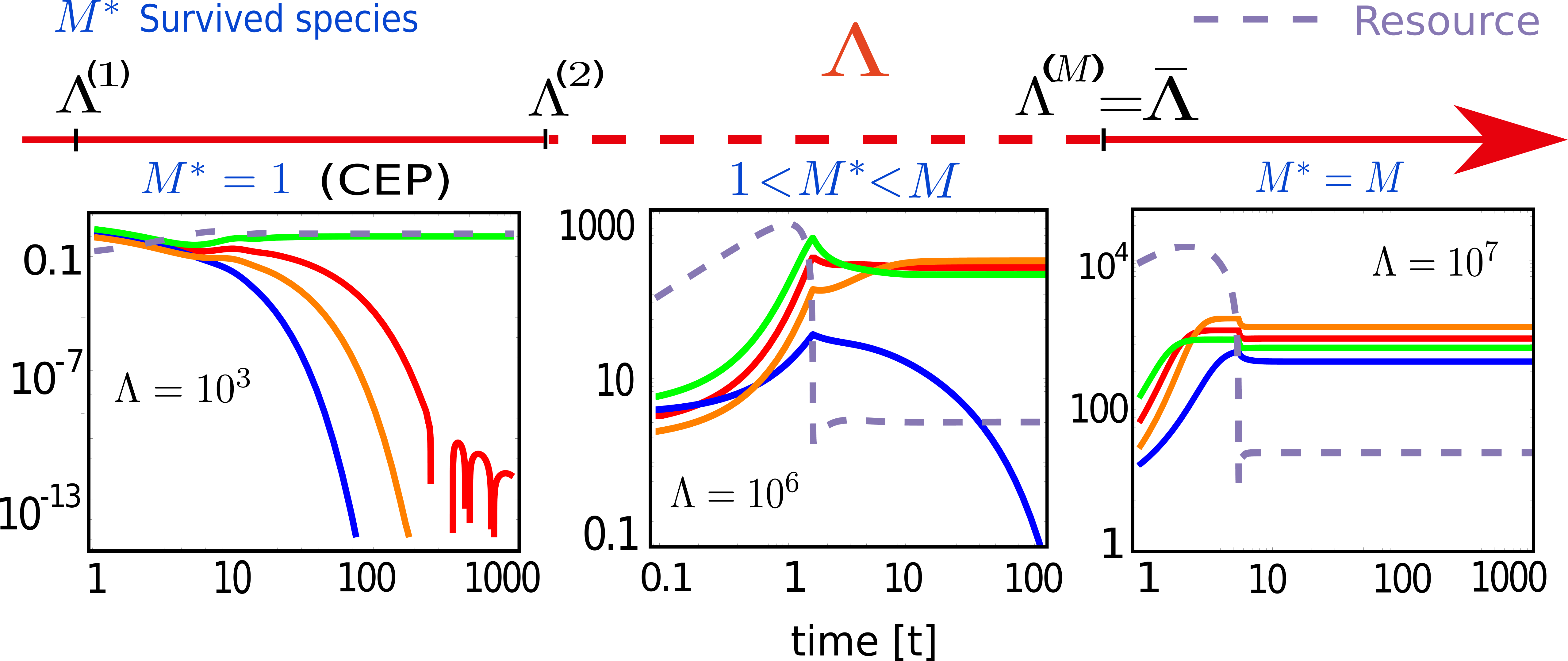}~
    \caption{Coexistence of species competing for 1 resource. Horizontal arrow indicates a schematic for $\Lambda$ such that $l$ number of species coexist if $\Lambda^{(l)}<\Lambda<\Lambda^{(l+1)}$ with $\Lambda^{(M+1)}=\infty$. We verify these results by numerically evolving dynamics \eqref{sp-eqn-3} and \eqref{res-eqn-3} for 4 species ($M=4$, solid lines) competing for one resource ($R=1$, dashed line). Clearly, for $\Lambda<\bar \Lambda$, the number of survived species $M^*$ is less than $M$. 
    See details of the parameters in \cite{SM}.}
    \label{fig:first}
\end{figure}
The numerical computation of the stationary values of species populations and resources from Eqs.~\eqref{sp-eqn-3} and \eqref{res-eqn-3} are showed in Fig.~\ref{fig:non-diag-E-species}. We have plotted the fraction of survived species as a function of the resource supply $\Lambda$, in the presence of one resource, by varying the ratio, $\epsilon_{\sigma \rho}/\epsilon_{\sigma \sigma}$, of inter- to intra-species interaction \footnote{If the distribution of the diagonal terms is $P_{\rm diag}(\epsilon)$ the distribution of the off-diagonal terms is $P_{\rm off-diag}(\epsilon)=a^{-1}P_{\rm diag}(\epsilon/a)$ where $a$ is the rescaling factor.}. As expected, when the ratio increases, the fraction of survived species decreases. All initial species survive when the inter-species interaction are relatively weaker, and an increasing number of species coexists (competing for one resource) when the resource supply $\Lambda$ is correspondingly larger (Fig.~\ref{fig:first}). It turns out that the limit $\epsilon_{\sigma \rho}\to 0$ for $(\sigma\neq \rho)$ is analytically tractable and, therefore, we will focus on this case in the following.

As the time progresses, we expect this system to reach a stationary state. If all species have survived (later we will discuss the case when a sub-set of them go extinct) at a large time $(n_\sigma^* >0 ~\forall~\sigma)$, then the following equations in the matrix form can be obtained by setting the left-hand side 
of Eqs.~\eqref{sp-eqn-3} and \eqref{res-eqn-3} equal to zero:
\begin{align}
\vec {\mathcal{N}}&=E^{-1} (Q G~\vec{\mathcal{U}}-\vec{\mathcal{B}}),\label{mat-1}\\
\mu(\vec{L}-\vec{\chi})&=G Q^T\vec{\mathcal{N}}\label{mat-2},
\end{align}
where $\vec{\mathcal{N}}=(n_1^*,n_2^*,\dots,n_M^*)^\top$, $\vec{\mathcal{B}}=(\beta_1,\beta_2,\dots,\beta_M)^\top$, $\vec{L}=(\Lambda_1,\Lambda_2,\dots,\Lambda_R)^\top$, $\vec{\chi}=(c^*_1,c^*_2,\dots,c^*_R)^\top$, $E=\text{diag}[\epsilon_1,\epsilon_2,\dots,\epsilon_M]$ (and it is invertible, where $\epsilon_{\sigma}\equiv \epsilon_{\sigma\sigma}$), $\vec{\mathcal{U}}=(1,1,\dots,1)^\top$ an $R$-component vector, $Q$ is a $M$$\times$$R$ matrix whose elements are the metabolic strategies ($[Q]_{\sigma i}$=$\alpha_{\sigma i}$), and  $G=\text{diag}[r_1(c_1^*),r_2(c_2^*),\dots,r_R(c_R^*)]$.
Substituting Eq.~\eqref{mat-1} in 
\eqref{mat-2} gives $R$ coupled equations:
\begin{align}
G Q^T E^{-1}QG \vec{\mathcal{U}}-G Q^T E^{-1} \vec{\mathcal{B}}=\mu(\vec{L}-\vec{\chi}),
\label{cond-11}
\end{align}
that can be solved for $r_i(c_i^*)$ as a function of the other
parameters. Further, the condition for all species to survive, 
using Eq.~\eqref{mat-1}, is  
 $(Q G~\vec{\mathcal{U}})_\sigma >\vec{\mathcal{B}}_\sigma~\forall~\sigma$,
~and it gives the coexistence region in the $R$-dimensional space whose axes are $r_1(c^*_1), r_2(c^*_2),\dots, r_R(c^*_R)$. Thus, a necessary condition for all initial species to coexist is that the solution of Eq.~\eqref{cond-11} lies within 
this coexistence region; otherwise, some of them go extinct. 

To illuminate the above result, we first consider a case when several species are competing for one resource (a discussion for higher number of resources is relegated to the Supplementary Material \cite{SM}). In this case, removing the immaterial index $i$, Eq.~\eqref{mat-1} becomes $n_\sigma^*=[\alpha_\sigma r(c^*)-\beta_\sigma]/\epsilon_\sigma~
\forall~\sigma$. Since $n_\sigma^*>0$, we find $r(c^*)>\beta_\sigma/\alpha_\sigma$. Moreover, we can write $r(c^*)>r(\bar c)\equiv\max_\sigma\{\beta_\sigma/\alpha_\sigma\}$, where $r(c^*)$ is the solution of Eq.~\eqref{cond-11}: $Ar^2(c^*)-Br(c^*)-\mu(\Lambda-c^*)=0$, in which the coefficients $A=\sum_\sigma \alpha_\sigma^2/\epsilon_\sigma$ and $B=\sum_\sigma \alpha_\sigma \beta_\sigma/\epsilon_\sigma$ carry the characteristic features of the species. Note that $r(c^*)\leq 1$, therefore, the metabolic strategies, in order to guarantee a coexistence of all species, should be greater than the death rates: $\alpha_\sigma>\beta_\sigma$.  Thus, for fixed parameters that characterize the species, i.e., $\{\alpha_\sigma,\beta_\sigma,\epsilon_\sigma\}$, coexistence of all species is achieved when tuning the resource supply rate by varying $\Lambda$ at a fixed $\mu$ in such a way that the condition $r(c^*)>r(\bar c)$ is satisfied. Such a critical value of $\Lambda$ is given by $\bar{\Lambda}=r(\bar c)[A r(\bar c)-B]/\mu+\bar c$. In Fig.~\ref{fig:first}, we consider an example of ecosystem having 4 species competing for 1 resource. Clearly, when $\Lambda>\bar\Lambda$, all initial species survive (shown by solid lines), while some of them go extinct in the contrasting case. 
\begin{figure}
  \includegraphics[width=\columnwidth]{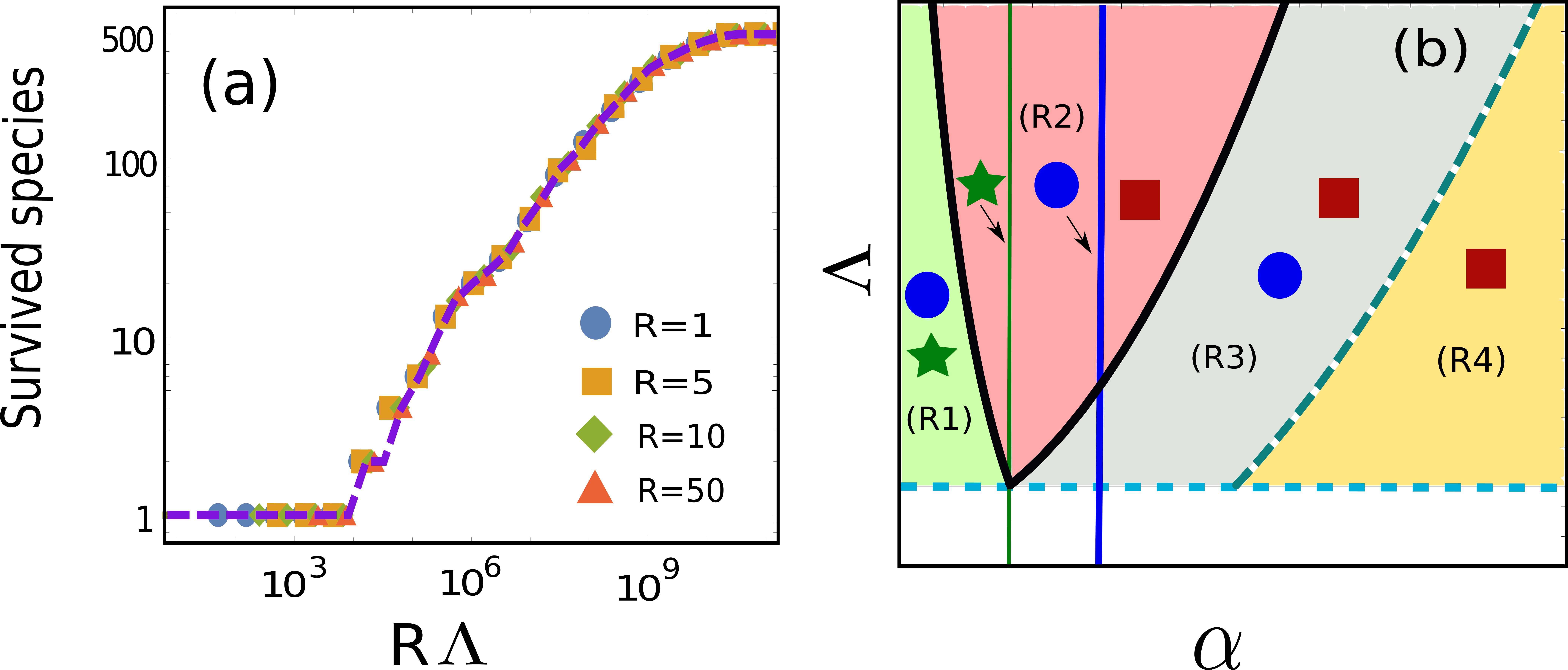}~~
    \caption{Survival of species with $\Lambda$. Panel (a): We compare the number of survived species obtained from numerical simulation (circles) of dynamics ~\eqref{sp-eqn-3} and \eqref{res-eqn-3} for $R=1$ with 
    theoretical prediction (dashed curve) (see text) for initial 500 species. Moreover, we compare the simulations results of $R=1$ with $R>1$ (square, diamond, and triangle). Clearly, all these curves collapse to each other when one rescales $\Lambda$ with $R$ (simulation details for $R>1$ are relegated to Supplementary Material \cite{SM}). Panel (b): Schematic of invasion by a third species (square) in a pool of two coexisting species (star and circle) competing for one resource. Two vertical lines correspond to metabolic strategies of two species (thicker for the fitter one, i.e., circle). We plot $\bar \Lambda$ (see main text), shown by a black solid curve [enclosing the region (R2)], that gives the critical resource supply for all of them to coexist as a function of metabolic strategy $\alpha$ of the invader for other fixed parameters. The dashed curve separating (R3) and (R4) corresponds to $\bar \Lambda$ for square and circle to coexist. The horizontal dashed line is the threshold $\Lambda$ above which circle and star coexist in the absence of invader. Four different regions (R1)--(R4) are shown depending on the survival of species.}
       \label{fig:inv}
\end{figure}

In the following, we discuss how many species (out of $M$ initial species) survive 
when competing for a single resource.
For simplicity, in what follows, we consider $\beta_\sigma=1$. Nevertheless, the analysis can also be done along the same line for generic $\beta_\sigma$'s. Since now each species is characterized by a set of parameters, one can define an array $\{\alpha_1,\epsilon_1;\alpha_2,\epsilon_2;\dots;\alpha_M,\epsilon_M\}$, where species are arranged according to decreasing metabolic strategies (see 
\cite{SM} for details). 
Further, we define two conditional sums: $A^{(c)}_l~=~\sum_{\sigma=1}^l\alpha_\sigma^2/\epsilon_\sigma,~B^{(c)}_l~=~\sum_{\sigma=1}^l \alpha_\sigma \beta_\sigma/\epsilon_\sigma$, 
for $1\leq l\leq M$. Next, similar to the case when all species coexist, we find $\Lambda^{(l)}=\tilde r(\bar c)\big[A_l^{(c)}\tilde r(\bar c)-B_l^{(c)}\big]/\mu+k\tilde r(\bar c)/[1-\tilde r(\bar c)]$, the critical supply for $l$ species in which $\tilde r(\bar c)=\max\{\alpha^{-1}_{\sigma}~\vert ~1\leq \sigma\leq l\}\equiv \alpha_l^{-1}$. Thus, if $\Lambda^{(l)}<\Lambda< \Lambda^{(l+1)}$ then $l$ species survive, $l=1,2,\dots , M$, where we have defined $\Lambda^{(M+1)}\equiv \infty$. In Fig.~\ref{fig:first}, we show both $\Lambda^{(1)}$ and $\Lambda^{(2)}$ (left markers) within which only the fittest species survives. We verify this result in Fig.~\ref{fig:first} by numerically evolving the dynamics ~\eqref{sp-eqn-3} and \eqref{res-eqn-3} for 4 species competing for one resource.

In Fig.~\ref{fig:inv}(a), we compare the theoretical prediction using $\Lambda^{(l)}$ for the number of coexisting species (dashed curve) with numerical simulations (circles) of Eqs.~\eqref{sp-eqn-3} and \eqref{res-eqn-3} for initial 500 species, and they have an excellent match. Moreover, we show the comparison for number of survived/coexisting species as $R$ increases  (see details in 
\cite{SM}). Interestingly, we find that the simulation data for $R>1$ collapse on the theoretical prediction for $R=1$ when the resource supply is scaled with the number of resources. This is because for the same number of species to coexist a less resource supply for each resource is required, as expected.

Thus, by simply taking into account the spatial effects through intra-species competition, we conclude that a new phase emerges  (in agreement with Fig.~\ref{fig:non-diag-E-species}) in addition to the phase where the number of coexisting species satisfies CEP. In the new phase, a finite fraction of all species coexists, regardless of the number of resources, and it saturates to $1$ at a finite resource rate.

By exploiting 
Fig.~\ref{fig:inv}(a), we now consider a pool of several coexisting species in the presence of one resource, and ask if a new species enters the community, when it will successfully invade or coexist with the other species?
To answer this question, for convenience, we consider a system of two coexisting species consuming one resource ($\Lambda$ above the dashed line) and a third one arrives [see a circle, a star, and a square (invader) in the schematic shown in Fig.~\ref{fig:inv}(b)]. 
Four different regimes [i.e., (R1)--(R4)] can be seen in Fig.~\ref{fig:inv}(b) depending on the various possibilities of surviving species.
These can be physically understood as follows. For a given $\Lambda$ (above the dashed line), only star and circle could survive  in region (R1) since $\Lambda$ is lower than the black solid curve. In (R2), $\Lambda$ is higher than the black curve, and therefore, we see coexistence. (R3) is the contrasting case to (R1), and finally, in (R4), the chosen $\Lambda$ is lower than the dashed curve indicating that invader can successfully invade the system (see
\cite{SM} for numerical simulation for each region.)

In addition to the prediction of the critical value of $\Lambda$ for the coexistence of a certain number of species and the related invasibility problem, our framework also allows us to determine the abundance distribution (SAD) of the surviving species, i.e., the probability density function $P(z)$ that a species has population size $z$.
Interestingly, we can obtain the exact $P(z)$ for the one resource case as well as a large number of resources (the calculations of the latter are presented in 
\cite{SM}). Nonetheless, for intermediate number of resources, $P(z)$ can be easily computed numerically.

Let us first consider the case in which all initial species survive. Herein, we find $n_\sigma^*=[\alpha_\sigma r(c^*)-1]/\epsilon_\sigma>0$ [see Eq.~\eqref{mat-1}]. Now, 
for a large number of species, one can think 
the parameters $\alpha_\sigma$ and $\epsilon_\sigma$ as random variables to incorporate the species variability and their differences. The stochasticity of these variables hinges on the way these are distributed among species. 
Let $Q_1(\alpha)$ and $Q_2(\epsilon)$, respectively, be the distributions for $\alpha$ and $\epsilon$ \footnote{Note that we have dropped the subscript $\sigma$ for convenience.}. Three different scenarios can be investigated: (1) $\alpha$ distributed as  a non-trivial $Q_1$ and $Q_2(\epsilon)=\delta(\epsilon-\hat{\epsilon})$, (2) $Q_1(\alpha)=\delta(\alpha-\hat{\alpha})$ and $\epsilon$ distributed as  a non-trivial $Q_2$, and  (3) both are non-trivial random variables. In the first scenario (1), the population distribution turns out to be \cite{SM} $P(z)=\hat{\epsilon} Q_1[(\hat{\epsilon} z+1)/r(c^*)]/r(c^*)$, in the second scenario (2), we find \cite{SM} $P(z)=[\hat{\alpha} r(c^*)-1]Q_2([\hat{\alpha} r(c^*)-1]/z)/z^2$, and finally in (3), we find the distribution (see 
details in \cite{SM}):
\begin{align}
P(z)&=\dfrac{[J_1(c,z)-J_1(d,z)]}{2z^2(q-p)(d-c)},
\label{pdf-1}
\end{align}
where $J_1(\kappa,z)=[(q^2-\kappa^2 z^2) \Theta(\kappa-p/z)+(q^2-p^2) \Theta(p/z-\kappa)]\Theta(q/z-\kappa)$ in which $p=a r(c^*)-1$ and $q=b r(c^*)-1$.  The above expression \eqref{pdf-1} is obtained for uniformly distributed $\alpha\in \mathcal{U}(a,b)$ and $\epsilon\in\mathcal{U}(c,d)$. Nonetheless, for any other distributions, the population distribution  $P(z)$ can also be computed \cite{SM}. 
Similarly, one can obtain the distribution of species' populations when only some of the initial ones survive (see \cite{SM}).

\begin{figure}
  \begin{center}
    \includegraphics[width=\columnwidth]{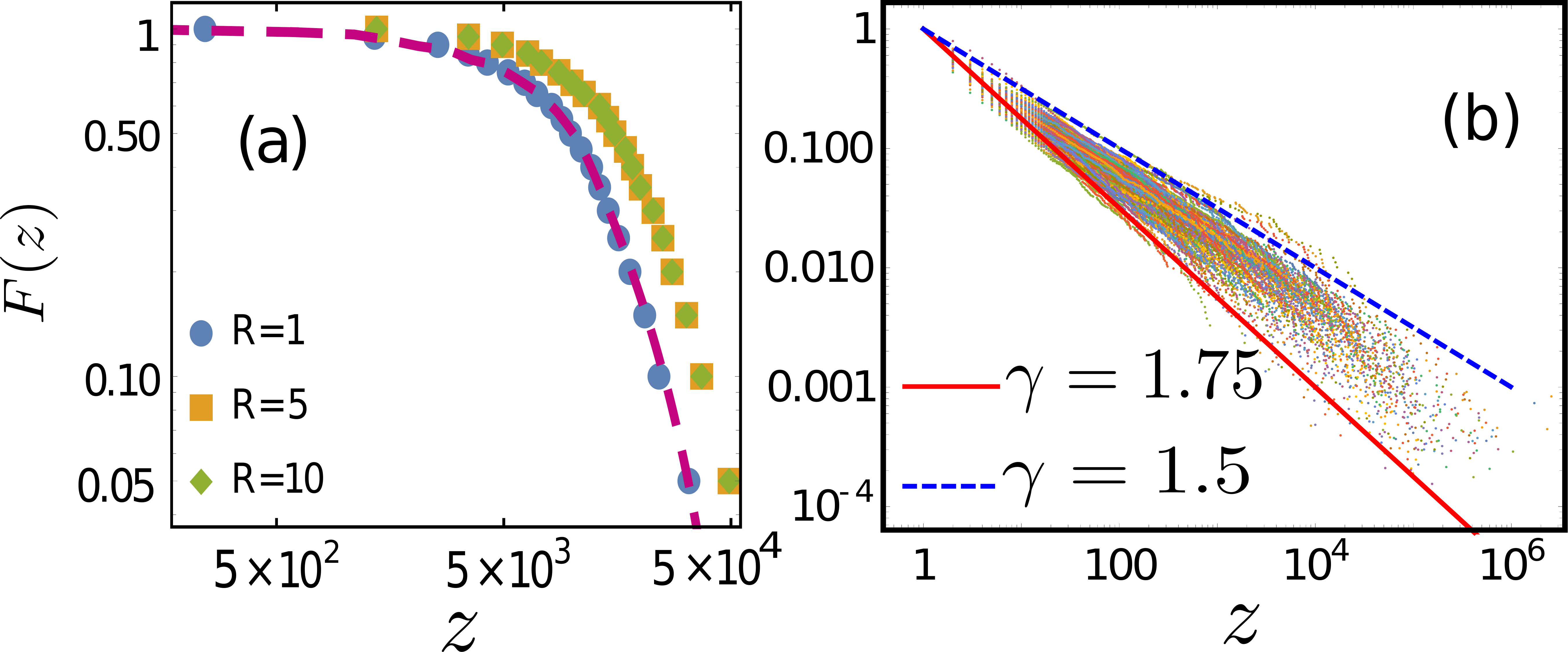}
    \caption{Complementary cumulative distribution function (CCDF) for populations of coexisting species. (a): Species CCDF, $F(z)$, is shown when all species coexist in the effective consumer-resource model. Circles are obtained by numerically integrating Eqs.~\eqref{sp-eqn-3} and \eqref{res-eqn-3} up to time $t=10^8$ for $M=500$ and one resource, while the dashed line is the analytical prediction of CCDF obtained from Eq.~\eqref{pdf-1}. Squares and diamonds, respectively, are the numerical simulations when the number of resources is $R=5$ or 10. All cases exhibit a similar trend. Further details are included in \cite{SM}. (b): Empirical CCDF for 134 surface seawater samples of microplankton obtained from the Tara ocean expedition \cite{de2015eukaryotic,data-1} (each color-code indicates one station). The solid lines are the theoretical predictions of the power-law decay $F(z)\propto z^{-\gamma +1}$ with exponents between $\gamma_l=1.5$ (blue dashed line) and $\gamma_u=1.75$ (red solid line) (see text and \cite{SM}).}
    \label{fig:4}
  \end{center}
\end{figure}

In Fig.~\ref{fig:4}(a), we plot the complementary cumulative distribution function (CCDF) $F(z)=\int_z^\infty~dy~P(y)$ of the population of the coexisting species. The exact prediction from Eq.~\eqref{pdf-1}  [dashed curves in Fig.~\ref{fig:4}(a)] is shown for the case of a single resource. Finally, we emphasize that in most of the cases (including a large number of resources \cite{SM}), the distribution $P(z)$ has power-law tails with exponent $-2$, i.e., $P(z)\sim z^{-2}$ as $z\to\infty$ \cite{SM}. The species population distribution has a similar power law behavior when the matrix $E$ is non-diagonal, but the off-diagonal entries are relatively smaller than the diagonal ones (see \cite{SM}). We stress that the power-law tail $z^{-2}$ for SAD is inevitable if $Q_2(\epsilon)$ is bounded in a neighbourhood of the origin.

As mentioned above, one of the celebrated examples where several species coexist even in the presence of a few resources is the ocean plankton \cite{paradox}. 
Recently, it has also been observed that species in plankton communities have a distribution of population sizes, as estimated by metagenomic studies, that decays as a steep power-law \cite{data-1,martin2020ocean}. Here we consider data on microplankton (20-180 $\mu m$ in body size) from the Tara ocean expedition \cite{de2015eukaryotic} (for more details, see \cite{data-1}, from which the data have been taken). In Fig.~\ref{fig:4}(b), we compare the empirical SADs (color-coded points) constructed from $134$ surface seawater samples distributed over all the oceans \cite{data-1} (each point corresponds to one station) to the SAD obtained from the stationary solution of our extended consumer-resource model (solid and dashed lines): $F(z)=\int_z^{\infty}P(x)dx\propto z^{-\gamma+1}$ for two ``extreme'' slopes $\gamma=\gamma_l=1.5$ and $\gamma=\gamma_u=1.75$. This was obtained from our model by considering a simple setting in which all species have the same $\alpha$, whereas the $\epsilon$-s follow a power-law distribution: $Q_2(\epsilon)\sim\epsilon^{\gamma-2}$, for small $\epsilon$, with $\gamma\in(1,2]$ (see \cite{SM}). We point out that a range of exponents between $\gamma_l$ and $\gamma_u$ can be reproduced by an appropriate tuning  of parameter $\gamma$ in $Q_2(\epsilon)$. We also consider datasets obtained from microbial communities \cite{grilli2020macroecological} and 
again find $F(z)$ is well captured by a power-law whose exponent is consistent with $Q_2(\epsilon)\sim\epsilon^{\gamma-2}$ (see \cite{SM}).

Finally, we remark that the shape of the SAD does not qualitatively change if one takes into account demographic stochasticity \cite{lande2003stochastic} in Eq.~\eqref{sp-eqn-3} in a phenomenological way \cite{SM}. Hence, these SAD patterns seem to indicate a different behaviour as compared to those obtained in a neutral theory framework \cite{reference-referee}.

In summary, we have extended the consumer-resource model by incorporating the inter- and intra-species contributions that arise by coarse-graining the spatial degrees of freedom and/or due to the presence of species specific pathogens. This model is able to predict how several species coexist even for a relatively small number of resources. Further, we obtained analytically the distribution of the population sizes for one and a large number of resources. Our results are supported by numerical simulations as well as the empirical SAD for plankton communities.\\

\textit{Acknowledgements.---} D.G., S.S., and A.M. are supported by ``Excellence Project 2018'' of the Cariparo foundation. S.S. acknowledges the Department of Physics and Astronomy from the University of Padova for funding through BIRD209912 project. S.G. acknowledges the support from Univeristy of Padova through the PhD fellowship within ``Bando Dottorati di Ricerca'' funded by the Cariparo foundation. All the authors thank the reviewers for their constructive criticisms that have contributed to improve the manuscript.


\begin{thebibliography}{10}

\bibitem{Book-1}
A.~Sher and M.C. Molles.
\newblock {\em Ecology: Concepts and Applications}.
\newblock McGraw-Hill Education, 2018.

\bibitem{thompson1994coevolutionary}
John~N Thompson.
\newblock {\em The coevolutionary process}.
\newblock University of Chicago press, 1994.

\bibitem{schluter2000ecology}
Dolph Schluter.
\newblock {\em The ecology of adaptive radiation}.
\newblock OUP Oxford, 2000.

\bibitem{chesson1997roles}
Peter Chesson and Nancy Huntly.
\newblock The roles of harsh and fluctuating conditions in the dynamics of
  ecological communities.
\newblock {\em The American Naturalist}, 150(5):519--553, 1997.

\bibitem{chesson2000mechanisms}
Peter Chesson.
\newblock Mechanisms of maintenance of species diversity.
\newblock {\em Annual review of Ecology and Systematics}, 31(1):343--366, 2000.

\bibitem{vincent1996trade}
TLS Vincent, D~Scheel, JS~Brown, and TL~Vincent.
\newblock Trade-offs and coexistence in consumer-resource models: it all
  depends on what and where you eat.
\newblock {\em The American Naturalist}, 148(6):1038--1058, 1996.

\bibitem{schmidt2010ecology}
Kenneth~A Schmidt, Sasha~RX Dall, and Jan~A Van~Gils.
\newblock The ecology of information: an overview on the ecological
  significance of making informed decisions.
\newblock {\em Oikos}, 119(2):304--316, 2010.

\bibitem{chase2003ecological}
Jonathan~M Chase and Mathew~A Leibold.
\newblock {\em Ecological niches: linking classical and contemporary
  approaches}.
\newblock University of Chicago Press, 2003.

\bibitem{paradox-2}
Francesco Carrara, Andrea Giometto, Mathew Seymour, Andrea Rinaldo, and Florian
  Altermatt.
\newblock Inferring species interactions in ecological communities: a
  comparison of methods at different levels of complexity.
\newblock {\em Methods in Ecology and Evolution}, 6(8):895--906, 2015.

\bibitem{Zelezniak6449}
Aleksej Zelezniak, Sergej Andrejev, Olga Ponomarova, Daniel~R. Mende, Peer
  Bork, and Kiran~Raosaheb Patil.
\newblock Metabolic dependencies drive species co-occurrence in diverse
  microbial communities.
\newblock {\em Proceedings of the National Academy of Sciences},
  112(20):6449--6454, 2015.

\bibitem{Goldford469}
Joshua~E. Goldford, Nanxi Lu, Djordje Baji{\'c}, Sylvie Estrela, Mikhail
  Tikhonov, Alicia Sanchez-Gorostiaga, Daniel Segr{\`e}, Pankaj Mehta, and
  Alvaro Sanchez.
\newblock Emergent simplicity in microbial community assembly.
\newblock {\em Science}, 361(6401):469--474, 2018.

\bibitem{paradox}
G.~E. Hutchinson.
\newblock The paradox of the plankton.
\newblock {\em The American Naturalist}, 95(882):137--145, 1961.

\bibitem{Arthur}
Robert~Mac Arthur.
\newblock Species packing, and what competition minimizes.
\newblock {\em Proceedings of the National Academy of Sciences},
  64(4):1369--1371, 1969.

\bibitem{Chesson-MA}
Peter Chesson.
\newblock Macarthur's consumer-resource model.
\newblock {\em Theoretical Population Biology}, 37(1):26 -- 38, 1990.

\bibitem{exclusion-1}
Garrett Hardin.
\newblock The competitive exclusion principle.
\newblock {\em Science}, 131(3409):1292--1297, 1960.

\bibitem{exclusion-2}
Robert MacArthur and Richard Levins.
\newblock Competition, habitat selection, and character displacement in a
  patchy environment.
\newblock {\em Proceedings of the National Academy of Sciences},
  51(6):1207--1210, 1964.

\bibitem{exclusion-3}
Simon~A. Levin.
\newblock Community equilibria and stability, and an extension of the
  competitive exclusion principle.
\newblock {\em The American Naturalist}, 104(939):413--423, 1970.

\bibitem{mcgehee1977some}
Richard McGehee and Robert~A Armstrong.
\newblock Some mathematical problems concerning the ecological principle of
  competitive exclusion.
\newblock {\em Journal of Differential Equations}, 23(1):30--52, 1977.

\bibitem{armstrong1980competitive}
Robert~A Armstrong and Richard McGehee.
\newblock Competitive exclusion.
\newblock {\em The American Naturalist}, 115(2):151--170, 1980.

\bibitem{britton1989aggregation}
NF~Britton.
\newblock Aggregation and the competitive exclusion principle.
\newblock {\em Journal of Theoretical Biology}, 136(1):57--66, 1989.

\bibitem{hening2020competitive}
Alexandru Hening and Dang~H Nguyen.
\newblock The competitive exclusion principle in stochastic environments.
\newblock {\em Journal of mathematical biology}, 80(5):1323--1351, 2020.

\bibitem{Wingreen}
Anna Posfai, Thibaud Taillefumier, and Ned~S. Wingreen.
\newblock Metabolic trade-offs promote diversity in a model ecosystem.
\newblock {\em Phys. Rev. Lett.}, 118:028103, Jan 2017.

\bibitem{Leo}
Leonardo Pacciani-Mori, Andrea Giometto, Samir Suweis, and Amos Maritan.
\newblock Dynamic metabolic adaptation can promote species coexistence in
  competitive microbial communities.
\newblock {\em PLOS Computational Biology}, 16(5):1--18, 05 2020.

\bibitem{Roy-2}
Shovonlal Roy and J.~Chattopadhyay.
\newblock The stability of ecosystems: A brief overview of the paradox of
  enrichment.
\newblock {\em Journal of Biosciences}, 32:421 -- 428, 2007.

\bibitem{caswell1976community}
Hal Caswell.
\newblock Community structure: a neutral model analysis.
\newblock {\em Ecological monographs}, 46(3):327--354, 1976.

\bibitem{hubbell1979tree}
Stephen~P Hubbell.
\newblock Tree dispersion, abundance, and diversity in a tropical dry forest.
\newblock {\em Science}, 203(4387):1299--1309, 1979.

\bibitem{hubbell2001unified}
Stephen~P Hubbell.
\newblock {\em The unified neutral theory of biodiversity and biogeography
  (MPB-32)}, volume~32.
\newblock Princeton University Press, 2001.

\bibitem{vallade2003analytical}
M~Vallade, , and B~Houchmandzadeh.
\newblock Analytical solution of a neutral model of biodiversity.
\newblock {\em Physical Review E}, 68(6):061902, 2003.

\bibitem{chave2004neutral}
Jer{\^o}me Chave.
\newblock Neutral theory and community ecology.
\newblock {\em Ecology letters}, 7(3):241--253, 2004.

\bibitem{alonso2006merits}
David Alonso, Rampal~S Etienne, and Alan~J McKane.
\newblock The merits of neutral theory.
\newblock {\em Trends in ecology \& evolution}, 21(8):451--457, 2006.

\bibitem{rosindell2011unified}
James Rosindell, Stephen~P Hubbell, and Rampal~S Etienne.
\newblock The unified neutral theory of biodiversity and biogeography at age
  ten.
\newblock {\em Trends in ecology \& evolution}, 26(7):340--348, 2011.

\bibitem{azaele2016statistical}
Sandro Azaele, Samir Suweis, Jacopo Grilli, Igor Volkov, Jayanth~R Banavar, and
  Amos Maritan.
\newblock Statistical mechanics of ecological systems: Neutral theory and
  beyond.
\newblock {\em Reviews of Modern Physics}, 88(3):035003, 2016.

\bibitem{hubbell1997unified}
Stephen~P Hubbell.
\newblock A unified theory of biogeography and relative species abundance and
  its application to tropical rain forests and coral reefs.
\newblock {\em Coral reefs}, 16(1):S9--S21, 1997.

\bibitem{volkov2003neutral}
Igor Volkov, Jayanth~R Banavar, Stephen~P Hubbell, and Amos Maritan.
\newblock Neutral theory and relative species abundance in ecology.
\newblock {\em Nature}, 424(6952):1035--1037, 2003.

\bibitem{azaele2006dynamical}
Sandro Azaele, Simone Pigolotti, Jayanth~R Banavar, and Amos Maritan.
\newblock Dynamical evolution of ecosystems.
\newblock {\em Nature}, 444(7121):926--928, 2006.

\bibitem{peruzzo2020spatial}
Fabio Peruzzo, Mauro Mobilia, and Sandro Azaele.
\newblock Spatial patterns emerging from a stochastic process near criticality.
\newblock {\em Physical Review X}, 10(1):011032, 2020.

\bibitem{bell2000distribution}
Graham Bell.
\newblock The distribution of abundance in neutral communities.
\newblock {\em The American Naturalist}, 155(5):606--617, 2000.

\bibitem{bell2001neutral}
Graham Bell.
\newblock Neutral macroecology.
\newblock {\em Science}, 293(5539):2413--2418, 2001.

\bibitem{data-1}
Enrico Ser-Giacomi, Lucie Zinger, Shruti Malviya, Colomban De~Vargas, Eric
  Karsenti, Chris Bowler, and Silvia De~Monte.
\newblock Ubiquitous abundance distribution of non-dominant plankton across the
  global ocean.
\newblock {\em Nature Ecology \& Evolution}, 2:1243--1249, 08 2018.

\bibitem{janzen1970herbivores}
Daniel~H Janzen.
\newblock Herbivores and the number of tree species in tropical forests.
\newblock {\em The American Naturalist}, 104(940):501--528, 1970.

\bibitem{connell1971role}
Joseph~H Connell.
\newblock On the role of natural enemies in preventing competitive exclusion in
  some marine animals and in rain forest trees.
\newblock {\em Dynamics of populations}, 298:312, 1971.

\bibitem{J-C}
Eugene~W. Schupp.
\newblock The janzen-connell model for tropical tree diversity: Population
  implications and the importance of spatial scale.
\newblock {\em The American Naturalist}, 140(3):526--530, 1992.

\bibitem{clark1984spacing}
Deborah~A Clark and David~B Clark.
\newblock Spacing dynamics of a tropical rain forest tree: evaluation of the
  janzen-connell model.
\newblock {\em The American Naturalist}, 124(6):769--788, 1984.

\bibitem{bagchi2010testing}
Robert Bagchi, Tom Swinfield, Rachel~E Gallery, Owen~T Lewis, Sofia Gripenberg,
  Lakshmi Narayan, and Robert~P Freckleton.
\newblock Testing the janzen-connell mechanism: pathogens cause
  overcompensating density dependence in a tropical tree.
\newblock {\em Ecology letters}, 13(10):1262--1269, 2010.

\bibitem{bagchi2014pathogens}
Robert Bagchi, Rachel~E Gallery, Sofia Gripenberg, Sarah~J Gurr, Lakshmi
  Narayan, Claire~E Addis, Robert~P Freckleton, and Owen~T Lewis.
\newblock Pathogens and insect herbivores drive rainforest plant diversity and
  composition.
\newblock {\em Nature}, 506(7486):85--88, 2014.

\bibitem{KELLER1971225}
Evelyn~F. Keller and Lee~A. Segel.
\newblock Model for chemotaxis.
\newblock {\em Journal of Theoretical Biology}, 30(2):225--234, 1971.

\bibitem{garcia2018bacterial}
  Leonor Garc{\'\i}a-Bayona and Laurie~E Comstock.
 \newblock Bacterial antagonism in host-associated microbial communities.
 \newblock {\em Science}, 361(6408), 2018. 


\bibitem{granato2019evolution}
 Elisa~T Granato, Thomas~A Meiller-Legrand, and Kevin~R Foster. 
\newblock The evolution and ecology of bacterial warfare.
\newblock {\em Current biology}, 29(11), 2019.



\bibitem{giometto2020antagonism}
 Andrea Giometto, David~R Nelson and Andrew~W Murray.
\newblock Antagonism between killer yeast strains as an experimental model for biological nucleation dynamics.
\newblock {\em bioRxiv}, 2020.



\bibitem{RG-1}
Kenneth~G. Wilson.
\newblock The renormalization group: Critical phenomena and the kondo problem.
\newblock {\em Rev. Mod. Phys.}, 47:773--840, Oct 1975.

\bibitem{RG-2}
Shang-keng Ma and Gene~F. Mazenko.
\newblock Critical dynamics of ferromagnets in
  $6\ensuremath{-}\ensuremath{\epsilon}$ dimensions: General discussion and
  detailed calculation.
\newblock {\em Phys. Rev. B}, 11:4077--4100, Jun 1975.



\bibitem{SM}
See Supplementary Material at [URL will be inserted by publisher] for
analytical derivations, additional mathematical details and further numerical
analysis, which contains Ref.~[58].



\bibitem{martin2020ocean}
Paula~Villa Martin, Ales Bucek, Tom Bourguignon, and Simone Pigolotti.
\newblock Ocean currents promote rare species diversity in protists.
\newblock {\em bioRxiv}, 2020.

\bibitem{de2015eukaryotic}
Colomban De~Vargas, St{\'e}phane Audic, Nicolas Henry, Johan Decelle,
  Fr{\'e}d{\'e}ric Mah{\'e}, Ramiro Logares, Enrique Lara, C{\'e}dric Berney,
  Noan Le~Bescot, Ian Probert, et~al.
\newblock Eukaryotic plankton diversity in the sunlit ocean.
\newblock {\em Science}, 348(6237), 2015.


\bibitem{grilli2020macroecological}
Jacopo Grilli.
\newblock Macroecological laws describe variation and diversity in microbial
  communities.
\newblock {\em Nature communications}, 11(1):1--11, 2020.


\bibitem{lande2003stochastic}
Russell Lande, Steinar Engen, Bernt-Erik Saether, et~al.
\newblock {\em Stochastic population dynamics in ecology and conservation}.
\newblock Oxford University Press on Demand, 2003.

\bibitem{reference-referee}
Rafael D'Andrea, Theo Gibbs, and James~P O’Dwyer.
\newblock Emergent neutrality in consumer-resource dynamics.
\newblock {\em PLoS computational biology}, 16(7):e1008102, 2020.

 \bibitem{May}
 {\color{black} Robert ~M May. 
 \newblock Will a large complex system be stable?
 \newblock {\em Nature}, 238(5346), 1972.
 }




\end{thebibliography}
\end{document}